# Optimal Design of a Molecular Recognizer: Molecular Recognition as a Bayesian Signal Detection Problem

Yonatan Savir and Tsvi Tlusty

*Abstract*—Numerous biological functions—such as enzymatic catalysis, the immune response system, and the DNA-protein regulatory network—rely on the ability of molecules to specifically recognize target molecules within a large pool of similar competitors in a noisy biochemical environment. Using the basic framework of signal detection theory, we treat the molecular recognition process as a signal detection problem and examine its overall performance. Thus, we evaluate the optimal properties of a molecular recognizer in the presence of competition and noise. Our analysis reveals that the optimal design undergoes a "phase transition" as the structural properties of the molecules and interaction energies between them vary. In one phase, the recognizer should be complementary in structure to its target (like a lock and a key), while in the other, conformational changes upon binding, which often accompany molecular recognition, enhance recognition quality. Using this framework, the abundance of conformational changes may be explained as a result of increasing the fitness of the recognizer. Furthermore, this analysis may be used in future design of artificial signal processing devices based on biomolecules.

## I. Introduction

SIGNAL processing in biological systems relies on the ability of bio-molecules to specifically recognize each other. Examples are antibodies targeting antigens, regulatory proteins binding to DNA and enzymes catalyzing their substrates. The *molecular recognizers* must locate and preferentially interact with their specific targets among a vast variety of molecules that are often structurally similar. This task is further complicated by the inherent noise in the biochemical environment, whose magnitude is comparable to that of the noncovalent binding interactions [1]–[3]. Optimization with respect to noise proves to be crucial in the design of biological information channels and especially of molecular codes [4], [5]. The task of the molecular recognizer is analogous to the task of a decision unit, which has to discern a specific signal within a collection of various noisy signals. This analogy motivates us to regard in the present work molecular recognition as a signal detection problem. Specifically, the framework of signal detection allows us to evaluate the properties of the optimal bio-recognizer, which must meet the severe requirements of recognition in a competitive noisy environment.

Many studies attempted to understand the remarkable specificity and efficiency of molecular recognition. It was realized early that recognizing molecules should be complementary in shape and, thus, discriminate against a competitor target that does not fit precisely to the recognizer binding sites, akin to matching lock and key. Later, however, it was found that the "native" forms of many recognizers and targets do not match exactly, and, therefore, they must deform in order to bind each other. Conformational changes upon binding have been observed in many bimolecular systems. For example in enzyme-substrate [6], antibody-antigen [7], [8] and other protein-protein complexes [9], [10], protein-DNA recognition [11], [12] and protein-RNA recognition [13]. The abundance of conformational changes raises the question of whether they occur due to biochemical constraints or whether they are perhaps the outcome of an evolutionary optimization of recognition processes. We address this question using signal detection theory, which provides a comprehensive framework for evaluating the optimality of recognition processes.

Let us consider a biological system that has to discriminate between two molecules, $A$ and $B$. The decision is made by binding to a recognizer molecule $a$. For example, $a$ may be an antigen and $A$ is a harmful pathogen that should be identified by $a$ while $B$ is a molecule normally found in the body and is similar to $A$. The typical recognition reaction, described by the Michaelis–Menten kinetics, consists of two basic steps [Fig. 1(a)]. In the first, reversible step, the recognizer $a$ binds $A$ or $B$ and produces the complexes $aA$ and $aB$. The formation of these complexes depends on the dissociation constants, $K_A = [a][A]/[aA]$, $K_B = [a][B]/[aB]$ ([ ] denotes concentrations), which relate the equilibrium concentrations of the complexes to the equilibrium concentrations of their components in the unbound form. In the second, irreversible step, these complexes initiate formation of a correct product at rate $R_A = \nu_A[aA]$ and formation of an incorrect product at rate $R_B = \nu_B[aB]$. The reaction rates $\nu_{A,B}$ have units of 1/time and are often referred to as the turnover numbers.

Previous studies focused mainly on the *specificity* of recognition, that is ratio between the correct and incorrect production rates $R_A/R_B$, and especially on the effect of recognizer-target structural match on specificity. It was argued, on a theoretical ground, that if the competing targets have the same structure, when bound to their recognizer, then conformational changes do not affect specificity [14], [15]. However, this conclusion is not valid if the competing targets have different structures [16], [17], or if the deformations are slow [12], [18]. Recently, we have studied how the flexibility of the molecules and the recognizer-target structural match affect specificity. We have shown that for competing targets with similar structure, optimal specificity is achieved by a recognizer that is complementary in structure to its main target (akin to "lock and key"). For competing targets with different structure, conformational mismatch (a recognizer which is slightly different from its main target) provides optimal specificity [17].

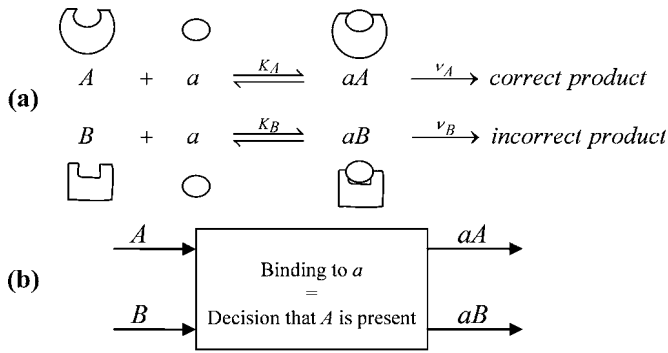

Fig. 1. Molecular binding as a detection problem. (a) Typical molecular recognition reaction. A recognizer $a$ can bind to two competing molecules $A$ and $B$ and, thus, initiate correct and incorrect actions. The reaction depends on the dissociation constants $K_{A,B}$ and on the production rates $\nu_{A,B}$ (see text). (b) Biological recognition system can be regarded as a detection problem where both the input-output signals and the decision unit are molecules. On the molecular level, the decision is carried out through binding of the recognizer $a$ to the "input" molecules, $A$ or $B$. The molecular binding, which is governed by the physical properties of the interacting molecules, dictates the decision quality.

However, there are other desirable properties of a recognition system besides specificity. The biological recognition system, as any other recognition system, should be specific but at the same time also robust to noise and efficient. For example, the recognizer needs to discriminate between different targets but still to recognize the main target at high enough probability. The recognizer also has to cope with noise arising from the inherent inaccuracies of biological systems, and, therefore, the recognition process has to be specific while being robust for noise. The tolerance of the system to errors is also crucial since, in some systems, misidentification may be harmless while in others it may be fatal. Besides the questions regarding the structural mismatch between the recognizer and target, i.e., whether conformational changes occur upon binding, the role of the flexibility of molecules should also be considered. We find that while rigid recognizers can execute highly specific recognition ("lock and key") they are very sensitive to errors. On the other hand, flexible recognizers may be more robust to noise but less specific. This calls for a more general measure for the quality of molecular recognition processes.

These considerations motivate our present work in which we formulate the molecular recognition process in terms of a Bayesian signal detection problem and, thus, introduce a comprehensive formalism for the design of an optimal bio-recognizer. Our analysis reveals two optimal "design phases." In one phase, the optimal recognizer and target have complementary shapes (a lock and a key). In the other phase, the optimal recognizer differs from its main target and, thus, conformational changes are beneficial. Unlike previous works, (including our specificity model [17]), this work shows that conformational changes may be beneficial even for targets with similar structures and reveals that the system undergoes a "phase transition" between the two designs as the flexibility of the molecules and the interaction energies vary. We evaluate the critical parameters of this transition and show that, in most cases, flexible biomolecules with a mismatch relative to their main target achieve optimal recognition performance.

## II. MOLECULAR RECOGNITION IN TERMS OF BAYESIAN DETECTION

Consider the biological system described in Fig. 1(a) that has to discriminate between two molecules $A$ and $B$. This discrimination process can be viewed as a signal detection problem. The input signals are the molecules $A$ and $B$, and the outcome is binding to the recognizer $a$, which is constrained by the physical interactions of the molecules [Fig. 1(b)]. In this simplified picture [Fig. 1(a)], the interacting molecules are regarded as rigid objects which may be complementary in structure. However, in the noisy biochemical environment, one expects both the recognizer and the targets to interconvert within an ensemble of many possible conformations. Such an ensemble may be the outcome, for example, of thermally induced distortions. In this case, all the conformations of the recognizer and the target may interact with each other and as a result a variety of complexes is formed [Fig. 2(a)]. These conformational fluctuations around the native state of the molecules are actually noise that should be added to the input signals $A$ and $B$ [Fig. 2(b)]. Such a fluctuating recognizer implies that the decision unit itself is prone to noise, that is the decision is made in a stochastic fashion.

Describing molecular recognition as a signal detection problem allows us to employ detection theory to evaluate the optimality of molecular recognition processes. A standard Bayesian decision rule is derived by minimizing the average Bayesian cost function, $C_b$, using posteriori probabilities [19]

$$C_b = \sum_{i,j} C_{ij} \cdot p_{h,j} \cdot p_d(i|j) \quad (1)$$

where $p_{h,j}$ is the probability for true hypothesis $j$ to occur, $p_d(i|j)$ is the probability for a decision or an outcome $i$ given true hypothesis $j$, and $C_{ij}$ is a cost assigned to a decision $i$ while the true hypothesis is $j$, which measures the consequences of each decision. This measure is often used in cases of simple hypotheses when the *a priori* probabilities for each hypothesis are known and obey $\sum_j p_{h,j} = 1$.

On the molecular level, the recognizer $a$, at an initial concentration $[a]_0$, is diffusing in an environment where both $A$ and $B$ may be present, at an initial concentrations $[A]_0$ and $[B]_0$, and may collide with each one of the targets. However, the probability that the recognizer collides with both of them simultaneously is practically zero; some fraction of the recognizers encounters $A$ molecules while the rest encounter $B$ molecules.

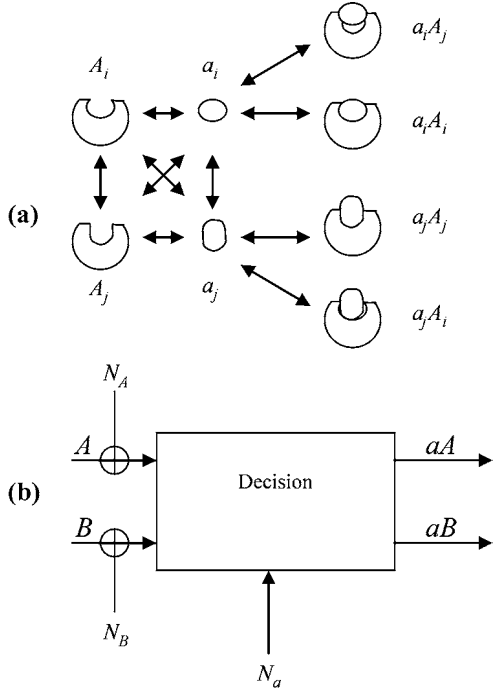

Fig. 2. General molecular recognition scheme. (a) Both the recognizer and the correct target are interconverting within an ensemble of conformations denoted by indices, $a_i$ and $A_i$, respectively. All the different conformations may interact and as a result a variety of complexes is formed. In a similar fashion, the various ligand conformations, $a_i$, may interact with competing target conformations $B_i$ and, thus, catalyze an incorrect product. (b) The flexibility of the target molecules is analogous to noise added to the input signals. The decision unit is also prone to noise since the recognizer $a$ is noisy, and, therefore, the decision is stochastic in nature.

The scenario is as if the system (an ensemble of recognizers) is exposed to $[A]_0$ at some probability $p_{h,A}$ and to $[B]_0$ at a probability $p_{h,B} = 1 - p_{h,A}$. The possible hypotheses are, therefore, exposing the system to initial concentrations $[A]_0$ or to initial concentration $[B]_0$.[1] The hypotheses probabilities $p_{h,A}$ and $p_{h,B}$ depend on the initial concentrations of the targets $[A]_0$ and $[B]_0$. For example, if the initial concentration of $B$ is zero, than the ensemble of recognizers encounters only $A$ and $p_{h,B} = 0$. If the initial concentrations are equal then $p_{h,A} = p_{h,B}$

The goal of the molecule $a$ is to recognize $A$ and induce a certain action. Hence, the possible decisions or outcomes are whether a function associated with the identification of $A$ is to be triggered or not. To initiate the desired function, a *functional* complex should be formed. The conditional decision probability, $p_{d,A/B} = \text{prob}$(formation of a functional complex $aA/aB$ | encounter with $[A]_0/[B]_0$), is the probability that a functional complex is formed given that the system has been exposed to some initial concentration. This probability is the product of the complex formation probability conditioned on encounter, $p_b$(complex formation|encounter), and the probability that the formed complex is functional, $p_f$, i.e., $p_{d,A/B} = p_{b,A/B} \cdot p_{f,A/B}$.

For example, in a possible immune system scenario, $a$ is an antigen, $A$ is a pathogen and $B$ is a harmless molecule. The possible outcomes are to trigger or not to trigger an immune response. Binding of $a$ to $A$ triggers correct immune response while binding of $a$ to $B$ leads to misidentification of $B$ as $A$ and the initiation of an incorrect immune response (Table I). In another example, $A$ and $B$ may be regulatory proteins, that is proteins that upon binding to an appropriate DNA sequence promote protein production, and $a$ is the DNA sequence corresponding to $A$. In this scenario, the possible outcomes are to produce or not produce the protein coded by $a$. In a similar way to the immune system scenario, binding of $a$ to $B$ may induce an incorrect response.

The decision to trigger a function, that is identification of $A$, is denoted by the sub-index $t$, while the decision not to trigger a function is denoted by the sub-index $nt$. The possible hypotheses, interaction with $[A]_0$ or interaction with $[B]_0$, are denoted by sub-indexes $A$ and $B$, respectively. A cost $C_{t/nt,A/B}$ is assigned for each possible scenario (Table I). It follows from (1) that the cost function is

$$C_b = c_A \cdot p_{h,A} \cdot p_{d,A} + c_B \cdot p_{h,B} \cdot p_{d,B} + \alpha \qquad (2)$$

where $c_A = C_{t,A} - C_{nt,A}$, $c_B = C_{t,B} - C_{nt,B}$ and $\alpha = p_{h,A} \cdot C_{t,A} + p_{h,B} \cdot C_{nt,B}$. Clearly, $\alpha$ does not depend on the binding probabilities or, in signal detection terminology, it is independent of how we assign points in the observation space. The goal of the optimization in the present case is to determine the structural properties that are optimal for detection. Formally, this means that one looks for the minima of $C_b$ as a function of these structural properties, which enter the problem only through the binding probabilities. Obviously, since $\alpha$ does not depend on the structural properties, it is irrelevant for the optimization problem and can be omitted hereafter.

In scenarios relevant to biological systems, the decision is usually facilitated by biochemical agents, which are produced by the complexes $aA$ or $aB$. Therefore, it is more natural to discuss the rate $\rho$ at which an existing complex induces product formation rather than the probability that this complex is functional, $p_f$. The rate $\rho$ is equal to the functionality probability times some reaction-rate, $\rho = \nu \cdot p_f$. Therefore, using rates instead of probabilities does not change the nature of the solution.

For clarity, we assume that the recognizer concentration is in excess relative to the target concentrations, $[a]_0 \gg [A]_0, [B]_0$, and, therefore [15]

$$p_{d,A} \cdot \nu_A = R_A = \nu_A \frac{[a]_0 [A]_0}{[a]_0 + K_A} \qquad (3)$$

and

$$p_{d,B} \cdot \nu_B = R_B = \nu_B \frac{[a]_0 [B]_0}{[a]_0 + K_B} \qquad (4)$$

where $K_{A,B}$ are the dissociation constants and $\nu_{A,B}$ are the turnover numbers discussed above. Using (3) and (4), the average cost analogous to (2) becomes

$$C = c_A \cdot p_{h,A} \cdot [A]_0 \frac{1}{1 + K_A/[a]_0} \cdot \nu_A$$
$$+ c_B \cdot p_{h,B} \cdot [B]_0 \frac{1}{1 + K_B/[a]_0} \cdot \nu_B \qquad (5)$$

---

[1]Although both targets are present together, this is a simple hypothesis scenario rather than a simultaneous hypotheses scenario, since the recognizer cannot interact with both targets simultaneously.

TABLE I
DECISION TABLE OF MOLECULAR RECOGNITION PROCESSES. DURING MOLECULAR RECOGNITION, THE DECISION IS MADE BY BINDING INTERACTIONS BETWEEN THE RECOGNIZER $a$ AND THE COMPETING TARGETS, $A$ AND $B$, WHICH LEAD TO THE FORMATION OF THE COMPLEXES $aA$ AND $aB$. THESE COMPLEXES MAY BE FUNCTIONAL AND PROMOTE SOME ACTION. THE RECOGNIZER $a$ IS DIFFUSING IN AN ENVIRONMENT WHERE BOTH $A$ AND $B$ MAY PRESENT. SINCE THE PROBABILITY THAT THE RECOGNIZER COLLIDES WITH BOTH OF THEM SIMULTANEOUSLY IS PRACTICALLY ZERO, THE SCENARIO IS AS IF THE SYSTEM (ENSEMBLE OF RECOGNIZERS) IS EXPOSED TO $[A]_0$ WITH SOME PROBABILITY $p_{h,A}$ AND TO $[B]_0$ WITH PROBABILITY $p_{h,B} = 1 - p_{h,A}$. THE ENCOUNTERS MAY BE FOLLOWED BY BINDING THAT INITIATES SOME ACTION. $p_{d,A/B}$ IS THE PROBABILITY TO FORM A FUNCTIONAL COMPLEX, $aA$ OR $aB$, GIVEN THE INITIAL CONCENTRATIONS $[A]_0$ OR $[B]_0$, RESPECTIVELY. THE TABLE DESCRIBES A MOLECULAR RECOGNITION DECISION TABLE USING A POSSIBLE IMMUNE SYSTEM SCENARIO AS AN EXAMPLE. BINDING OF AN ANTIGEN $a$ SHOULD INDICATE THE PRESENCE OF A PATHOGEN $A$. THEREFORE, BINDING OF $a$ TO A HARMLESS MOLECULE $B$ WILL RESULT IN A "FALSE ALARM." UNBINDING AFTER AN ENCOUNTER BETWEEN $a$ AND $A$ IS A "MISS." SIMILAR DECISIONS DESCRIBE A POSSIBLE DNA-PROTEIN REGULATION SCENARIO. IN THIS CASE, $A$ AND $B$ ARE REGULATORY PROTEINS WHILE $a$ REPRESENTS A SPECIFIC DNA SEQUENCE CORRESPONDING TO $A$. BINDING OF $a$ TO $A$ WILL RESULT IN FORMATION OF A CORRECT PROTEIN WHILE BINDING TO $B$ CAUSES MISIDENTIFICATION OF $A$ AND RESULTS IN UNDESIRED PROTEIN PRODUCTION

| Hypothesis / Decision | Interaction with a pathogen $A$ at initial concentration $[A]_0$; (A)  $p_{h,A}$ | Interaction with a harmless molecule $B$ at initial concentration $[B]_0$; (B)  $p_{h,B} = 1 - p_{h,A}$ |
|---|---|---|
| Trigger immune response; (t) | **Correct Decision**  $a$ binds $A$ and a correct functional complex $aA$ is formed. Correct Immune response is triggered.  $p_{d,A} = prob(\text{functional } aA \mid [A_0])$  $C_{t,A}$ | **Incorrect Decision ('False alarm')**  $a$ binds $B$ and an incorrect functional complex $aB$ is formed. Incorrect immune response due to misidentification of $A$ as $B$.  $p_{d,B} = prob(\text{functional } aB \mid [B_0])$  $C_{t,B}$ |
| Do not trigger immune response; (nt) | **Incorrect Decision ('Miss')**  $a$ interacts with $A$ but no functional complex is formed.  $1 - p_{d,A}$  $C_{nt,A}$ | **Correct Decision**  $a$ interacts with $B$ but no functional complex is formed. No misidentification of $B$ as $A$.  $1 - p_{d,B}$  $C_{nt,B}$ |

This form of the cost will be used to examine the optimization of biological recognition systems. In its current form, (5) has a simple biochemical meaning, it is a linear sum of the correct production rate and the incorrect production rate. We show below that the nature of physical interaction, including the flexibility of the recognizer and its mismatch relative to the correct target, affects the dissociation constants and, thus, the average cost.

III. DEPENDENCE OF THE COST ON STRUCTURAL PARAMETERS OF THE MOLECULES

The average cost function $C$ (5) that was derived in the last section provides a measure for the quality of a molecular recognition system. This cost function depends on the structural parameters of the participating molecules such as their flexibility and structural match. Next, in order to evaluate the optimal design of molecules, we calculate the dependence of the cost on these parameters.

Molecular recognition is a complex process, which involves searching for the molecular target and recognizing it via specific binding interaction. Binding requires the alignment of the active sites and conformational changes of the participating molecules. This complicated dynamics can be simplified into energetic considerations by using the reasonable assumption that recognition takes place close to thermodynamic equilibrium. In essence, molecular recognition is governed by the interplay between the interaction energy gained from the alignment of the binding sites and the elastic energy required to deform the molecules to align. Since the participants molecules interconvert within ensembles of conformations, various complexes may be formed. In some of these complexes the recognizer and the target may have a complementary structure and, therefore, may be functional. Accounting for all possible complexes and functionalities determines the overall cost.

Motivated by deformation spectra measurements and various numerical studies [22], we treat the molecules as elastic networks and take into account only the lowest elastic mode. Modeling proteins as elastic networks was previously applied to study large amplitude [20] and thermal fluctuations [21] of proteins, and to predict deformations and domain motion upon binding [22], [23]. Taking into account only the lowest elastic mode is a vast simplification of the many degrees of freedom that are required to describe the details of a conformational change. Nevertheless, this simplified model still captures the essence of the energy tradeoff.

To study the binding-deformation tradeoff, consider a flexible recognizer with a binding domain on which $N$ binding sites are distributed (Fig. 3). This recognizer interacts with an elastic target with a binding domain on which $N$ complementary binding sites are distributed. Binding is specific, i.e., each binding site on the recognizer site can interact only with its counterpart binding site on the target. This interaction occurs when the corresponding binding sites are close. Therefore, binding energy $E_b = N\varepsilon$ is gained if the binding domains are aligned and the binding sites can interact. However, to gain this binding energy the molecules must deform and deformation of the molecules from their native state costs elastic energy. We consider here only the elastic deformations in which the binding domain of the ligand is deformed uniformly, which are the lowest-energy modes. The distortion energy is, therefore, quadratic in some geometrical mismatch $d$ between the molecules, $E = (1/2)kd^2$, where $k$ is an effective spring constant and $d$ is the structural *mismatch* between the native structures, which may be length difference in the case of simple extension or angle difference in the case of simple bending (Fig. 3). The dissociation constant of a flexible target with a spring constant $k_t$ and a flexible recognizer with a spring constant $k_r$ is (Appendix A)

$$K = \frac{[\text{recognizer}][\text{target}]}{[\text{recognizer} - \text{target complex}]}$$
$$= \frac{Z_k}{V}\sqrt{\frac{2\pi}{\beta \bar{k} g^2}} e^{-\beta(E_b - \frac{\bar{k}}{2}d^2)} \quad (6)$$

where $\bar{k} = k_r k_t/(k_r + k_t)$ is the harmonic mean of the recognizer and target spring constants, $Z_k$ is kinetic partition function of the molecules (which does not depend on $d$ and $k_{r/t}$) and $V$ is the system volume. $\beta = 1/k_B T$ where $k_B$ is Boltzmann constant, $T$ is the system temperature and $k_B T$ is the typical thermal energy. $g$ is the typical scale of the binding interaction and is usually of order of few angstroms. As the binding energy $E_b$ increases, $K$ decreases and the molecules tend to form complexes. If $\bar{k}d^2$ increases, that is more energy is required for deformation, $K$ increases and the molecules tend to dissociate.

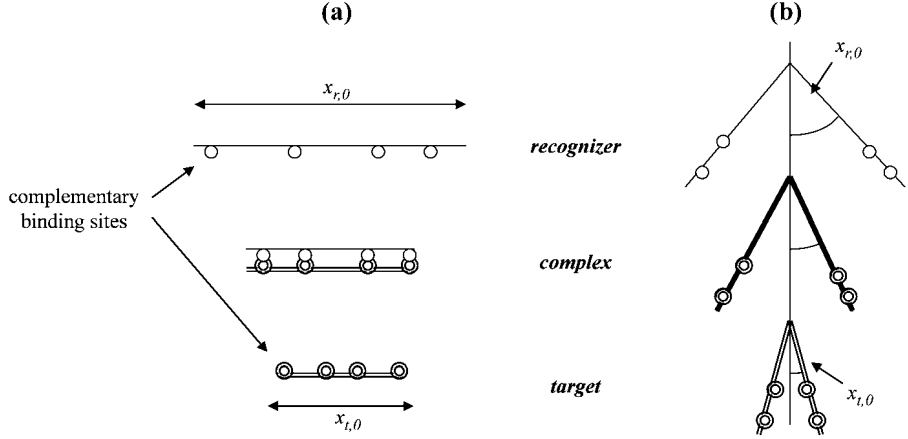

Fig. 3. Deformation of molecules upon binding in the case of (a) uniform stretching or shrinking and (b) uniform bending. On both the recognizer and the target there are $N$ specific binding sites. Both the recognizer and the target are fluctuating around native structures, $x_{r,0}$ and $x_{t,0}$, respectively, which may be the length the binding domain (a) or the angle of the binding domain (b). The dissociation constant (6) depends exponentially on the structural *mismatch* between the native structures, $d = x_{r,0} - x_{t,0}$.

The quality of a recognition process depends on two main characteristics of the participant molecules, their chemical affinity and the conformational match between them. To discuss the conformational effect, we consider a main or "correct" target $A$ and an "incorrect" competitor $B$ that differ in structure; for example, their binding domains may be of different lengths, $x_{A,0} \neq x_{B,0}$. Chemical affinity is taken into account by assuming that the competing target $B$ has only $N - m$ interacting binding sites while the main one has $N$, and, therefore, the binding energies, $E_{b,A} = N\varepsilon$, $E_{b,B} = (N - m)\varepsilon$, are different. The average cost (5) can now be expressed in terms of the structural mismatch $d$ and the spring constants of the molecules, $k_A$, $k_B$, as

$$C = \frac{-1}{1 + \frac{s}{\sqrt{\bar{k}_A}} e^{-\beta E_{b,A}} e^{\beta \frac{\bar{k}_A}{2} d^2}} + \frac{\chi}{1 + \frac{s}{\sqrt{\bar{k}_B}} e^{-\beta E_{b,B}} e^{\beta \frac{\bar{k}_B}{2}(d+\Delta)^2}} \quad (7)$$

where $\Delta = x_{t,A} - x_{t,B}$ is the structural difference between the competing targets and $s = (Z_k/V[a]_0)\sqrt{2\pi/\beta g^2}$ is a measure for the kinetic phase space of the molecules. The parameter $\chi = -(c_B \cdot p_{h,B} \cdot [B]_0 \cdot \nu_B)/(c_A \cdot p_{h,A} \cdot [A]_0 \cdot \nu_A)$ represents the *tolerance* of the system. As $\chi$ increases, the system is less tolerant to errors. For example, increase of $\chi$ may result from increasing the initial concentration of the incorrect target, $B$, relative to the initial concentration of the correct one, $A$, from raising the cost of incorrect decision relative to the correct one or from increasing the rate of incorrect product formation relative to the correct one. Similarly, as $\chi$ is smaller, the system is more tolerant to errors. By minimizing the cost (7), we evaluate below the optimal decision rule which depends on the mismatch and flexibility of the participant molecules. As a consequence, we find the characteristics of the optimal recognizer.

## IV. OPTIMAL DESIGN OF MOLECULAR RECOGNIZERS

The optimal flexibility of the molecules and the optimal mismatch of the recognizer relative to its main targets can be obtained by minimizing the average cost function $C$ (7). This optimal design depends on the properties of the competing targets, which may differ in structure and flexibility. Within the signal detection analogy, this means that the input signals may vary, for example they may have different frequency or amplitude, and that the noise affecting them may also vary. Optimizing the recognizer flexibility is equivalent to optimizing the amount of noise or stochasticity in the detection unit.

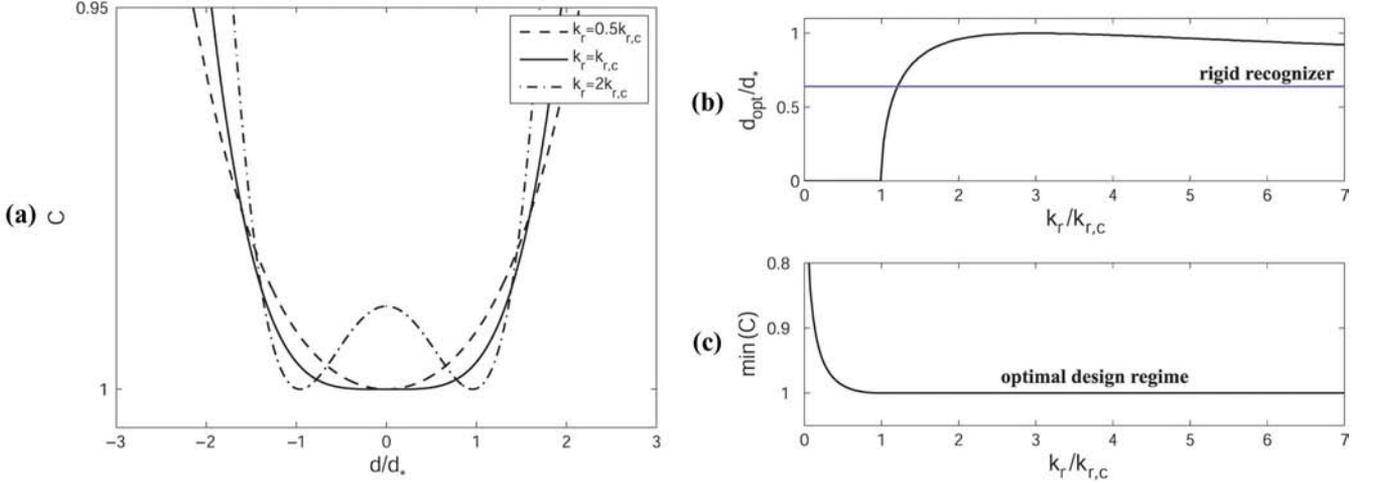

Fig. 4. Optimal mismatch for competing targets with similar structure. (a) Average cost for different values of recognizer flexibility, $k_r$. Below a critical flexibility, $k_{r,c}$, the cost $C$ has only one minimum at zero mismatch. Above the critical flexibility new minima emerge at a nonzero mismatch. (b) Optimal mismatch, $d = x_r - x_{t,A}$, as a function of recognizer flexibility, $k_r$. Very flexible recognizers, $k_r < k_{r,c}$, are easily distorted by both targets and, therefore, for those recognizers, the optimal mismatch is zero. However, if the recognizer is more rigid, $k_r > k_{r,c}$, the correct target deforms it more easily than the incorrect one and the optimal mismatch is nonzero. As the recognizer becomes more rigid, it is not easily deformed even by the correct target and the optimal mismatch tends to $d_{\text{rigid}}$ as $k_r^{-1/2}$. (c) The average cost gets its minimal value above $k_{r,c}$ and remains constant. Therefore, for noisy targets with similar structure, that is flexible targets with $k_t > k_c$, the optimal design is a recognizer with nonzero mismatch, even if the recognizer is rigid.

Our analysis exhibits two optimal design "phases." In one phase, the optimal recognizer and the target have complementary structures (a lock and a key), while in the other phase the optimal recognizer differs from its main target, and, thus, conformational changes are beneficial. The system undergoes a "phase transition" as the flexibility of the molecules and interaction energies between them vary. In the following section, we first analyze the case in which the competing targets have similar shape and flexibility, that is the targets are prone to similar noise. Then, we consider the case in which the targets are different both in structure and flexibility and analyze the effect of these differences on the optimal design. Finally, we estimate the typical biological parameters to evaluate the optimal design for biomolecules.

### A. Competing Targets With Similar Structural Properties

Consider the case in which the targets have a similar structure, $\Delta = x_{t,A} - x_{t,B} = 0$, and similar flexibility i.e their spring constants are equal, $k_A = k_B = k_t$. The targets differ in their chemical affinities, more binding energy is gained from binding of the recognizer to the correct target than from binding to the incorrect one, $|E_{b,A}| > |E_{b,B}|$.

First, we consider the symmetric case in which the tolerance parameter is $\chi = 1$. In this case, 1) the costs assigned to the correct and incorrect decisions are equal in magnitude and of opposite sign $c_A = -c_B$, 2) the initial concentrations of the correct target and the incorrect one are equal, $[A]_0 = [B]_0$, and 3) the functionalities of the competing complexes are the same, $\nu_{aA} = \nu_{aB}$.

As shown in Fig. 4, for a very flexible recognizer (low $k_r$), the cost $C$ exhibits a minimum at zero mismatch, $d = 0$, which means that the recognizer and targets should have complementary structures. However, above a certain critical spring constant, $k_{r,c}$, new minima emerge and the system undergoes a second-order phase transition as the mismatch which minimizes the average cost changes smoothly from zero to a nonzero value. Thus, *conformational changes become beneficial*. The optimal recognizer should differ from its main target, i.e the "key" should not be complementary to its "lock" but slightly different.

The critical spring constant of the recognizer at this transition is

$$k_{r,c} = \frac{k_c k_t}{k_t - k_c} \qquad (8)$$

where

$$k_c = s^2 e^{-\beta(E_{b,A} + E_{b,B})}. \qquad (9)$$

Above $k_{r,c}$, the optimal mismatch, $d_{\text{opt}}$, is

$$d_{\text{opt}} = \sqrt{\frac{\log(\bar{k}/k_c)}{\beta \bar{k}}} \qquad (10)$$

where $\bar{k} = k_r k_t / (k_r + k_t)$. For a very flexible recognizer, $k_r < k_{r,c}$, the distortion energy is much smaller than the binding energy of both targets and, thus, introducing a mismatch does not provide any advantage. This is also the case if $k_t < k_c$, that is very flexible targets. However, above some critical flexibility, $k_t > k_c$, $k_r > k_{r,c}$, the deformation energy is higher and, thus, deformation can occur upon binding to the correct target, while upon binding to the incorrect it is not likely to occur. Therefore, it is beneficial to introduce a slightly deformed recognizer and the optimal mismatch has a nonzero value.

Above the transition point, the optimal mismatch increases and reaches a maximal value of $d_* = 1/\sqrt{\beta e k_c}$ at $k_r = e k_{r,c}$. At this point, the deformation energy is equal to $k_B T$, the typical thermal energy [Fig. 4(b)]. Above this maximum point, the optimal mismatch decreases as the recognizer spring constant increases. As the recognizer becomes more rigid, more energy is required for its deformation. If the deformation energy exceeds $k_B T$, the thermal energy, even the binding energy of the correct target may not be sufficient for deforming the recognizer, and, thus, the optimal mismatch slightly decreases. Yet, the optimal mismatch for a rigid recognizer, $k_r \to \infty$, is still nonzero. As the recognizer becomes more rigid, the optimal mismatch tends to $d_{\text{rigid}} = (1/\sqrt{\beta k_t})\sqrt{\log(k_t/k_c)}$ as $k_r^{-1/2}$. The average cost function decreases as $k_r$ increases and becomes constant above $k_{r,c}$ [Fig. 4(c)]. Therefore, this super-critical regime, in which the optimal mismatch is nonzero, is also the *optimal design regime*.

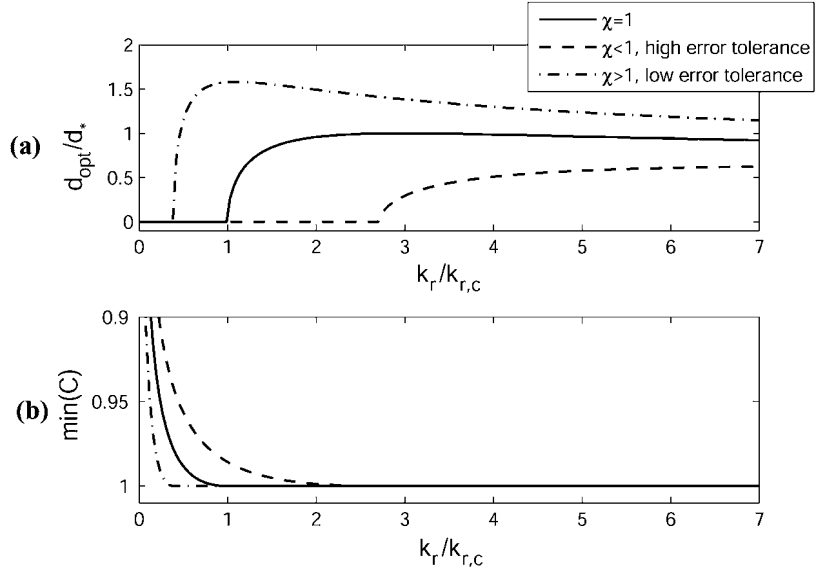

Fig. 5. Systems with different tolerance. The optimal mismatch (a) and minimal cost (b), for different values of the tolerance parameter $\chi$. When $\chi > 1$, avoiding wrong decisions is of higher priority than making the correct ones, that is the tolerance of the systems to errors is lower. Therefore, the critical flexibility is lower and a mismatch is introduced even for relatively more flexible recognizers. For $\chi < 1$, the error tolerance is high and the priority is the formation of correct products. In this case, the critical flexibility is increased (see text).

When the tolerance is asymmetric, $\chi \neq 1$, the system still undergoes a phase transition but the values of the critical parameters depend on $\chi$ (Fig. 5). When $\chi > 1$ the tolerance of the systems to errors is reduced (relative to $\chi = 1$) and avoiding a wrong decision is of higher priority than making the correct one. As a result, the critical spring constant is lower and mismatch should be introduced even for relatively flexible recognizers. For $\chi < 1$ the tolerance of the system is higher and the priority is the formation of correct product. Therefore, mismatch is introduced only at higher flexibilities.

### B. Competing Targets With Different Structural Properties

So far we have discussed targets with similar structural properties. However, differences in the flexibilities of the competing targets, that is, input signals that are prone to different noise levels, and structures also affect the optimal design.

*1) Targets With Different Flexibilities, $k_A \neq k_B$:* Molecules with different spring constants fluctuate differently. Due to thermal fluctuations, both targets are interconverting within ensembles of conformations. These conformations have a different mismatch $d$ relative to the recognizer and are distributed according to Boltzmann distribution, $p(d) \sim exp(-\frac{k}{2}d^2)$, where $k$ is the effective spring constant of the molecule. Hence, the distribution of the conformations is a Gaussian centered at $d = 0$ with a variance $\sigma \sim 1/\sqrt{k}$. Rigid molecules fluctuate less than flexible ones and, therefore, are less noisy.

When the incorrect target is more flexible, that is noisier, than the correct one, $k_{t,A} > k_{t,B}$, the critical flexibility is higher than in the case where $k_{t,A} = k_{t,B}$ [Fig. 6(a)]. Moreover, the average cost is minimal at one specific value of flexibility, $k_r \approx k_c$ [Fig. 6(b)]. Since the incorrect target fluctuates more, the ensemble of incorrect target conformations is more "smeared" than the correct ensemble. A flexible recognizer with zero mismatch can "sample" many correct conformations while sampling only few incorrect ones. Therefore, in this case, the *optimal design is a noisy recognizer*. In other words, a decision unit which is subject to noise will perform better than a deterministic decision unit.

In the case where the correct target is noisier, $k_{t,A} < k_{t,B}$, the critical flexibility is lower than in the case where $k_{t,A} = k_{t,B}$. Above the critical flexibility, the cost $C$ continues to decrease as a function of recognizer flexibility. In this case, the ensemble of correct target conformations is more spread than the incorrect ensemble. Thus, the *optimal design is a rigid recognizer with a nonzero mismatch* relative to the main target.

*2) Targets With Different Structure, $\Delta = x_{t,A} - x_{t,B} \neq 0$:* The nature of optimal design also depends on the structural differences between the targets. If the targets differ slightly, $\Delta \approx d_*$, as in the scenario of two similar targets, there is a phase transition at $k_r \cong k_{r,c}$ and new minima emerge (Fig. 7). However, the optimal mismatch is nonzero for any value of $k_r$. In the symmetric case [Fig. 4(c)], the minimal average cost value drops sharply after the transition point and remains constant, indicating that the super-critical regime is the optimal design regime. In the nonsymmetric case, the minimal average cost value also decreases sharply at the transition point but continues to decrease slowly. Therefore, the super-critical regime is again better than the subcritical regime but the global optimal design is of a rigid recognizer with nonzero mismatch. Unlike the symmetric case, the minima of the average cost are not symmetric [Figs. 4(a) and 7(a)], and, therefore, the optimal recognizer should slightly differ from its correct target but differ even more from the incorrect one. Targets which differ much, $\Delta \gg d_*$, are not actual competitors and therefore, in this case, the optimal design is a rigid recognizer with perfect complementarity as expected.

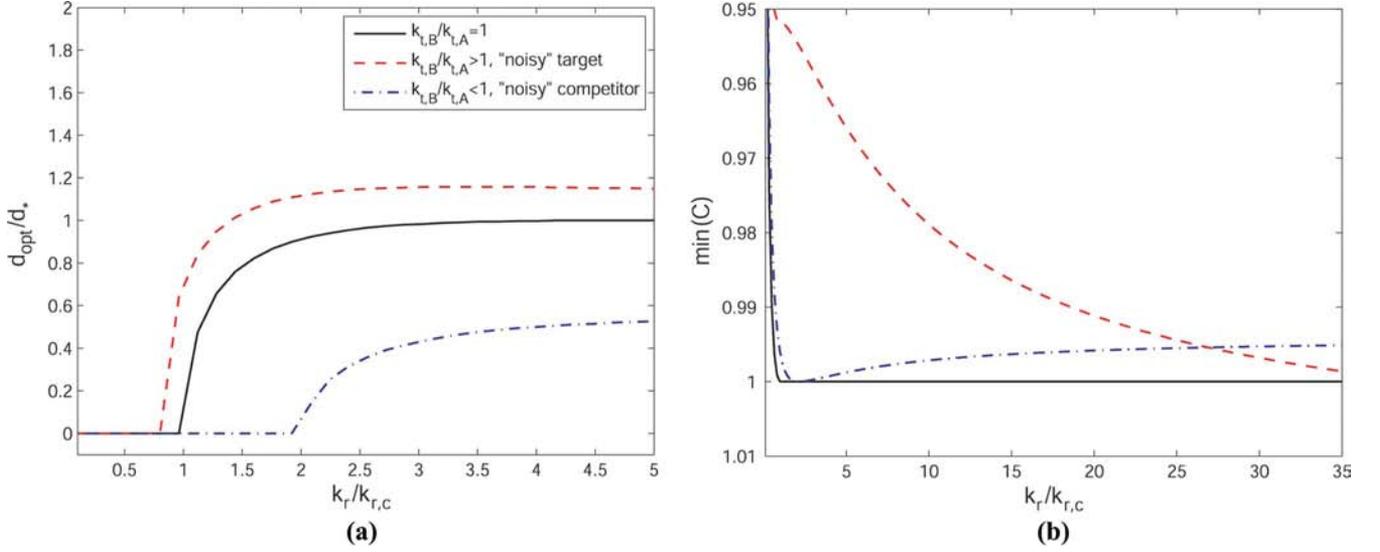

Fig. 6. Competing targets with different flexibility. The optimal mismatch (a) and minimal cost (b), in the case of targets of different flexibilities. If the competing targets have different flexibilities, $k_A \neq k_B$, that is the signals are prone to different magnitudes of noise, the optimal design depends on the relative magnitude of noise. If the incorrect target is noisier, i.e more flexible $k_{t,A} > k_{t,B}$, the average cost function is minimal for a recognizer with a spring constant which is about $k_{r,c}$ and the optimal design is a flexible recognizer with almost zero mismatch. If the correct target is noisier, $k_{t,A} < k_{t,B}$, then the optimal design is a rigid recognizer with a finite mismatch.

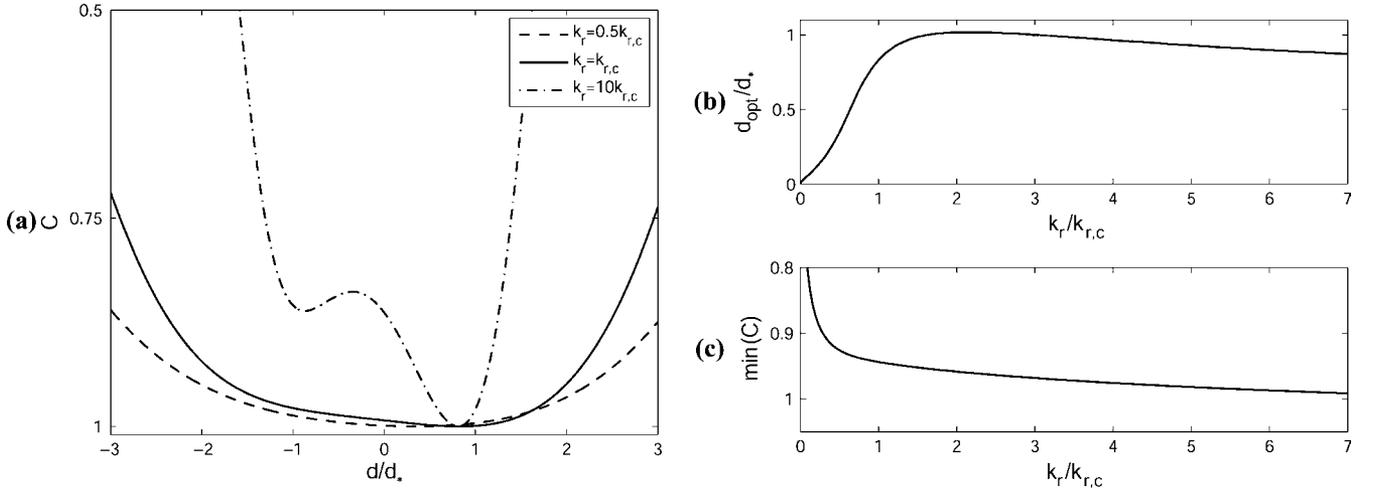

Fig. 7. Competing targets with different structure, $\Delta = (1/4)d_*$. (a) The average cost for different values of recognizer flexibility. Like in the case of similar targets (Fig. 4), below the critical spring constant, $k_{r,c}$, $C$ has only one minimum and above the critical spring constant new minima emerge. However, in this case, the optimal mismatch is nonzero for all values of $k_r$ and the super-critical minima are not symmetric. The recognizer should slightly differ from the main target but differ even more from the incorrect one. (b) Optimal mismatch as a function of recognizer flexibility. Very flexible recognizer, $k_r < k_{r,c}$, is easily distorted by both targets, and, therefore, the optimal mismatch is small. As the recognizer flexibility approaches its critical value the optimal mismatch increases "smoothly." As the recognizer becomes more rigid, it is not easily deformed even by the correct target and the optimal mismatch tends to its asymptotic value as $k_r^{-1/2}$, similar to the symmetric case. (c) The minimal value of the average cost drops sharply near the critical spring constant, $k_{r,c}$, so the super-critical regime is again the favored design regime. Unlike the symmetric case, after the transition, the minimal average cost slowly decreases and as a result, a rigid recognizer with nonzero mismatch is the global optimal design.

### C. Design Phases of Molecular Recognizer

In the previous section we discussed the optimal design phases, which depend on the structural parameters of the recognizers. Using the relation between the dissociation constant (6) and the critical spring constant (9), the typical realistic flexibility regime of optimal molecular recognizers can be evaluated. To find these flexibilities, we examine the ratio between the recognizer spring constant $k_r$ and the critical recognizer spring constant $k_{r,c}$ which is given by

$$\frac{k_r}{k_{r,c}} = (r+1)\frac{[a]_0^2}{K_A K_B e^{-2\beta E_d}} - r \qquad (11)$$

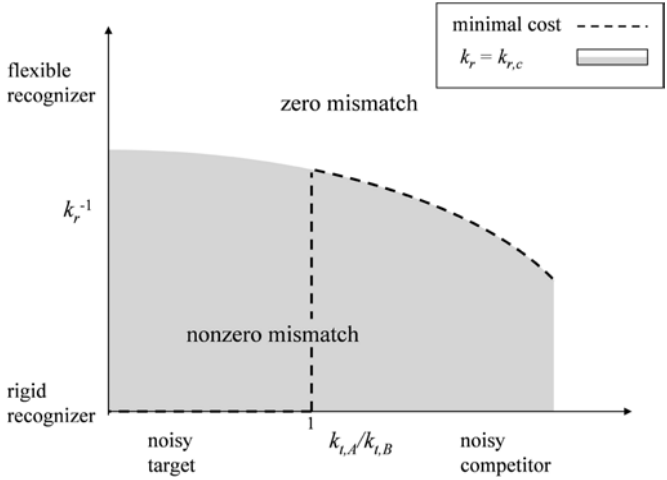

Fig. 8. Optimal design phases. The optimal mismatch for a recognizer depends on the flexibility of the recognizer and on the relative noise in the competing targets. The dashed line is the design with minimal cost value. The line $k_r = k_{r,c}$ is the transition line between a design with zero mismatch and a nonzero mismatch design. For a noisy competitor, the optimal design is along the $k_r = k_{r,c}$ line, that is a flexible recognizer with a zero mismatch. For a noisy target, the optimal design is a rigid target with a mismatch. Typical flexibilities values of biomolecules corresponds to $k_r \gtrsim k_{r,c}$ and thus both designs, with or without mismatch, may be advantageous.

where $r$ is the ratio between the recognizer and target spring constants, $r = k_r/k_t$ and $E_d = (k_t/2)d^2$ is the distortion energy. $K_{A,B}$ are the dissociation constants of the correct and incorrect targets, respectively. $[a]_0$ is the recognizer initial concentration.

Recognizer with mismatch is beneficial if $k_r > k_{r,c}$. Substituting typical biological parameters: $K_{A,B} \sim 10^{-4} - 10^{-16}$ M, $E_d \sim 10 k_B T$, $[a]_0 \sim 10^{-6} - 10^{-9}$ M, [24], [25], yields that $k_r \gtrsim k_{r,c}$. It follows from (11) that a mismatch is introduced if the dissociation constants of the targets satisfy $K_A K_B < ([a]_0^2/e^{-2\beta E_d}) \sim 10^{-3} - 10^{-9}$ M$^2$. Therefore, since $k_r \gtrsim k_{r,c}$, the bio-molecules are in a regime where both designs, with or without mismatch, may be advantageous.

## V. CONCLUSION

By treating molecular recognition processes as a signal detection problem, we evaluated the optimal design of a molecular recognizer in a noisy biochemical environment. The "phases" of optimal design are presented in Fig. 8. The optimal mismatch between the recognizer and the target depends on whether the recognizer flexibility is above or below the critical one. Typical structural parameters and interactions of biomolecules indicate that the recognizer spring constant is of the same magnitude or larger than the critical spring constant. Thus, both design strategies, with or without mismatch, are relevant for molecular recognition scenarios.

We also evaluated the flexibility of optimal recognizers. If the competitor is noisier than the correct target, the optimal design is a recognizer with critical flexibility and a zero mismatch. Such a flexible recognizer corresponds to stochastic detection unit. This implies that, in this case, a noisy recognizer samples the target more efficiently than a deterministic, rigid, recognizer. If the correct target is noisier, the optimal design is a rigid recognizer with a nonzero mismatch. In all cases, the significant decline in the average cost is at the transition between a zero mismatch design and a nonzero mismatch design. Hence, the regime in which the optimal mismatch is nonzero is the optimal design regime.

Although our model for distortion upon binding simplifies the real situation by considering only the lowest elastic modes, the approach of analyzing molecular recognition as a decision system may shed light on the nature of biological recognition systems. More realistic and empirical energy functions may be introduced into this framework to evaluate the optimal design. The result that, in most cases, conformational changes provide optimal recognition, may explain their abundance in nature as a mechanism that increases the fitness of the recognizer [26]. Besides explaining observed processes, this kind of formalism may be used in the design of future synthetic biological recognition systems.

## APPENDIX A
## DISSOCIATION CONSTANT EVALUATION

Both the recognizer and the target are interconverting within ensembles of conformations due to thermal fluctuations. Thus, the structure of those molecules is fluctuating around some native state structure according to Boltzmann distribution. The native state can be characterized by the length of the binding domain, angle, area or any other geometrical coordinate (Fig. 3). Since we consider only elastic deformations, the distortion energy is $(k/2)(x - x_0)^2$ where $k$ is an effective spring constant and $x_0$ is the native state configuration. Since all reactions besides product formations are assumed to be in equilibrium, we may apply the law of mass action

$$K_A = [a][A]/[aA] = \frac{1}{V}\frac{Z_a Z_A}{Z_{aA}} \quad (12)$$

where $V$ is the system volume and $Z_a$, $Z_A$, and $Z_{aA}$ are the recognizer, target, and complex partition functions, respectively. The partition function calculation is straightforward

$$K_A = \frac{Z_k}{V}$$
$$\times \frac{\int_0^\infty e^{-\beta \frac{k_t}{2}(x_t - x_{t,0})^2} dx_t \int_0^\infty e^{-\beta \frac{k_r}{2}(x_r - x_{r,0})^2} dx_r}{\int_0^\infty \int_0^\infty e^{-\beta\left(\frac{k_t}{2}(x_t-x_{t,0})^2 + \frac{k_r}{2}(x_r-x_{r,0})^2 - E_b\right)} \delta(x_t - x_r) dx_t dx_r} \quad (13)$$

$Z_k$ is the kinetic partition function, $E_b$ is the energy gain due to binding, subscript $r$ indicates recognizer, and subscript $t$ indicates target. The delta function in the denominator ensures that the complex is assembled out of complementary molecules. Under the reasonable assumption that $k_r^{1/2} x_{r,0}, k_t^{1/2} x_{t,0} \gg 1$ (assumption made mainly for clarity) performing integration yields

$$K_A = \frac{Z_k}{V}\sqrt{\frac{2\pi}{\beta \bar{k} g^2}} e^{-\beta(E_b - \frac{\bar{k}}{2}d^2)} \quad (14)$$

where $\bar{k} = k_r k_t/(k_r + k_t)$ and the normalization factor $g$ reflects the fact that the ensemble is continuous. $g$ is proportional to the typical interaction length of the binding sites.

ACKNOWLEDGMENT

The authors would like to thank U. Alon and D. Tawfik for valuable discussions.

## REFERENCES

[1] L. Pauling, *Science*, vol. 92, pp. 77–79, 1940.
[2] W. Kauzmann, "Some factors in the interpretation of protein denaturation," *Adv. Protein Chem.*, vol. 14, pp. 1–63, 1959.
[3] C. Tanford, "Interfacial free energy and the hydrophobic effect," *Proc. Nat. Acad. Sci. USA*, vol. 76, pp. 4175–4176, 1979.
[4] T. Tlusty, "A simple model for the evolution of molecular codes driven by the interplay of accuracy, diversity and cost," *Phys. Biol.*, vol. 5, p. 016001, 2008.
[5] T. Tlusty, "Rate-distortion scenario for the emergence and evolution of noisy molecular codes," *Phys. Rev. Lett.*, vol. 100, pp. 048101–048104, 2008.
[6] G. G. Hammes, "Multiple conformational changes in enzyme catalysis," *Biochemistry*, vol. 41, pp. 8221–8228, 2002.
[7] R. Jimenez, G. Salazar, K. K. Baldridge, and F. E. Romesberg, "Flexibility and molecular recognition in the immune system," *Proc. Nat. Acad. USA*, vol. 100, pp. 92–97, 2003.
[8] L. C. James and D. S. Tawfik, "Structure and kinetics of a transient antibody binding intermediate reveal a kinetic discrimination mechanism in antigen recognition," *Proc. Nat. Acad. USA*, vol. 102, pp. 12730–12735, 2005.
[9] S. Jones and J. M. Thornton, "Principles of protein-protein interactions," *Proc. Nat. Acad. USA*, vol. 93, pp. 13–20, 1996.
[10] C. S. Goh, D. Milburn, and M. Gerstein, "Conformational changes associated with protein-protein interactions," *Curr. Opin. Struct. Biol.*, vol. 14, pp. 104–109, 2004.
[11] C. O. Pabo and R. T. Sauer, "Protein-DNA recognition," *Annu. Rev. Biochem.*, vol. 53, pp. 293–321, 1984.
[12] K. A. Johnson, "Conformational coupling in DNA polymerase fidelity," *Annu. Rev. Biochem.*, vol. 62, pp. 685–713, 1993, [15].
[13] J. R. Williamson, "Induced fit in RNA-protein recognition," *Nat. Struct. Biol.*, vol. 7, pp. 834–837, 2000.
[14] A. R. Fersht, "Catalysis, binding and enzyme-substrate complementarity," *Proc. Roy. Soc. Lond. B, Biol. Sci.*, vol. 187, pp. 397–407, 1974.
[15] A. R. Fersht, *Structure and Mechanism in Protein Science: A Guide to Enzyme Catalysis and Protein Folding*. New York: Freeman, 1998.
[16] C. B. Post and W. J. Ray, Jr., "Reexamination of induced fit as a determinant of substrate specificity in enzymatic reactions," *Biochemistry*, vol. 34, pp. 15881–15885, 1995.
[17] Y. Savir and T. Tlusty, "Conformational proofreading: The impact of conformational changes on the specificity of molecular recognition," *PLoS ONE*, vol. 2, p. e468, 2007.
[18] D. Herschlag, "The role of induced fit and conformational changes of enzymes in specificity and catalysis," *Bioorg. Chem.*, vol. 16, pp. 62–96, 1988.
[19] C. W. Helstrom, *Elements of Signal Detection and Estimation*. Englewood Cliffs, NJ: Prentice-Hall, 1995.
[20] M. M. Tirion, "Large amplitude elastic motions in proteins from a single-parameter, atomic analysis," *Phys. Rev.Lett.*, vol. 77, pp. 1905–1908, 1996.
[21] I. Bahar, A. R. Atilgan, and B. Erman, "Direct evaluation of thermal fluctuations in proteins using a single-parameter harmonic potential," *Fold Des.*, vol. 2, pp. 173–181, 1997.
[22] F. Tama and Y. H. Sanejouand, "Conformational change of proteins arising from normal mode calculations," *Protein Eng.*, vol. 14, pp. 1–6, 2001.
[23] D. Tobi and I. Bahar, "Structural changes involved in protein binding correlate with intrinsic motions of proteins in the unbound state," *Proc. Nat. Acad. USA*, vol. 102, pp. 18908–18913, 2005.
[24] C. Kleanthous, *Protein-Protein Recognition*. Oxford, U.K.: Oxford Univ. Press, 2000.
[25] L. Stryer, *Biochemistry*, 3rd ed. New York: Freeman, 1988.
[26] Y. Savir and T. Tlusty, "DNA extension induced by RecA during homologous recombination as a possible conformational proofreading mechanism," to be published.

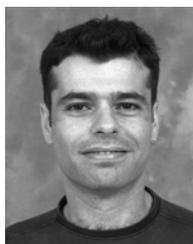

**Yonatan Savir** received the B.Sc. degree in electrical engineering and the B.A. degree in physics, both *summa cum laude*, from The Technion—Israel Institute of Technology, Haifa, Israel, and the M.Sc. degree in physics from the Weizmann Institute of Science, Rehovot, Israel, in 2004 and 2006, respectively. He is currently pursuing the Ph.D. degree in physics at the Department of Physics of Complex Systems, Weizmann Institute of Science.

His research interests include design principles and evolution of biological system and molecular information processing.

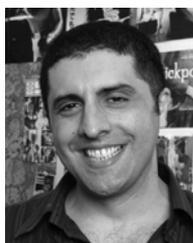

**Tsvi Tlusty** is a Senior Researcher with the Department of Physics of Complex Systems at the Weizmann Institute of Science, Rehovot, Israel. His research interests include molecular information-processing systems and their evolution.